\documentstyle[aps,prl,twocolumn,floats,psfig]{revtex}
\input{epsf}
\begin{document}
\draft
\twocolumn[\hsize\textwidth\columnwidth\hsize\csname@twocolumnfalse\endcsname

\title{New Universality Class at the Superconductor--Insulator Transition}

\author{Karl-Heinz Wagenblast$^{a)}$, Anne van Otterlo$^{b)}$, 
  Gerd Sch\"on$^{a)}$, and Gergely T. Zim\'anyi$^{c)}$}
\address{a) Institut f\"ur Theoretische Festk\"orperphysik,  
  Universit\"at Karlsruhe, D-76128 Karlsruhe, Germany}
\address{b) Theoretische Physik, ETH--H\"onggerberg, CH-8093 Z\"urich, 
  Switzerland}
\address{c) Department of Physics, University of California, 
  Davis, CA 95616, USA}
\date{24 August 1996}
\maketitle

\begin{abstract}
  We study dynamic properties of thin films near the
  superconductor - insulator transition. We formulate the problem in a
  phase representation.  The key new feature of our model is the
  assumption of a {\it local} ohmic dissipative mechanism. Coarse
  graining leads to a Ginzburg-Landau description, with non-ohmic
  dynamics for the order parameter.  For strong enough damping a new
  universality class is observed. It is characterized by a 
  {\it non-universal} d.c.\ conductivity, and a damping dependent
  dynamical critical exponent. The formulation also provides
  a description of the magnetic field-tuned transition. Several
  microscopic mechanisms are proposed as the origin of the
  dissipation.
\end{abstract}

\pacs{PACS numbers: 74.76.-w, 74.50.+r}
]

Granular superconductors and Josephson junction arrays behave
similarly in many respects. This is because the key degrees of freedom
are thought to be exclusively bosonic in nature, the Cooper pairs of
the underlying electronic problem. Quantum phase transitions are
present in the ordered arrays, when superconductivity gives way to a
gapped (Mott-) insulator. This is a direct consequence of the
uncertainity relation between phase and charge degrees of freedom
\cite{doni}. When disorder is present, an additional insulating phase
may appear, exhibiting glassy behaviour~\cite{fish3,davis}. In two
dimensions, early experimental studies reported the conductivity
$\sigma$, to assume a universal value~\cite{orr}. A theoretical
explanation of this universality was proposed based on scaling
considerations~\cite{fish1}.  However, subsequent measurements observed
critical resistivities, differing as much as
tenfold~\cite{liu,valles,yazd}. The main motivation of our work is to
develop an understanding of this apparent non-universality.

In this paper we develop a Ginzburg-Landau-Wilson (GLW) formulation
for the superconducting order-parameter. We emphasize the importance
of electronic excitations to account for the experiments. 
Our main result is the observation of a new universality class at the
superconductor - insulator transition. The conductivity at criticality
is found to be {\em non-universal}, and the dynamical critical
exponent depends on model parameters for strong enough dissipation.
The {\it non-ohmic} dynamics of the order parameter leads to a
power-law behaviour of the conductivity at low frequencies in the
insulating phase.  The magnetic field-tuned transition will be also discussed.

Our starting point is an array of superconducting islands or
grains connected by Josephson junctions. The $i^{\text{th}}$ 
grain is characterized by the phase $\varphi_i$ of its superconducting
order-parameter. It is also important to include the low energy 
electronic degrees of freedom. Integrating out the electrons leads
to a dissipative dynamics for the local phases. (For a more detailed
physical motivation see below.) One arrives at the Euclidean action 
\begin{eqnarray}\nonumber
  S[\varphi] &=& \int_{0}^{\beta}d\tau \Big[
  \sum_{i}\frac{\dot{\varphi}_i(\tau)^2}{2U}
  -J\sum_{\langle ij\rangle}\cos[\varphi_i(\tau)-\varphi_j(\tau)]\Big]\\
  &+&\frac{1}{2}\int_{0}^{\beta} d\tau d\tau'\sum_{i} 
  \alpha(\tau-\tau')\left[\varphi_i(\tau)-\varphi_i(\tau')\right]^2\,.
\label{Sjja}
\end{eqnarray}
The smallness of the grains manifests itself in a charging energy $U$
which is the energy cost of transferring an extra Cooper pair on the
island.  The Josephson coupling between neighboring islands is denoted
by $J$.  If the low lying electronic excitations have a finite density
of states, then the damping will be ohmic; its Fourier transform is
thus given by $\alpha(\omega)=\alpha_0|\omega_\mu|/2\pi$.  The
locality of the damping is reflected in the fact that the dissipative term
in Eq.~(\ref{Sjja}) couples the phase in a single island at different
times. Additional non-local charging or dissipative terms do not change
qualitatively the results and for simplicity will be dropped. 

A Hubbard-Stratonovich transformation is employed to decouple
the Josephson term \cite{doni}. This introduces the
complex order-parameter field $\psi$ such that its expectation value 
is proportional to that of $\exp(i\varphi)$. 
This yields the GLW action which in two dimensions takes the form
\begin{eqnarray}\nonumber
  F[\bar{\psi},\psi]&=&
  \int d^2r\;d\tau d\tau'\bar{\psi}(r,\tau)\bigg\{ \frac{1}{2J}\left(
      1-\frac{\vec{\nabla}^2}{4}\right)\delta(\tau-\tau')\\
    & &-g(\tau-\tau')\bigg\} \psi(r,\tau')
   +\kappa\int d^2r\; d\tau|\psi(r,\tau)|^{4}\;.
\label{free}
\end{eqnarray}
The correlation function $g$ is given as:
$g(\tau)=\langle\exp\{i\varphi_i(\tau)-i\varphi_i(0)\}\rangle_{0}$,
where $\langle\cdots\rangle_0$ is an expectation value taken with
respect to the single site Gaussian part of the action of
Eq.~(\ref{Sjja}).  In the presence of local damping $g(\tau)$ decays
algebraically in time $(\sim\tau^{-2/\alpha_0})$.  The Fourier
transform for small frequencies reads
\begin{equation}
  g(\omega_{\mu})=\frac{1}{2J}\left\{g_0-\eta\left|\omega_{\mu}\right|^{s}
    -\zeta \omega_{\mu}^2\right\};
  \hspace{3mm} s=\frac{2}{\alpha_0}-1\;,
\end{equation}
where $\eta= -4J\Gamma(-s)\cos(s\pi/2)(4/U\alpha_0)^{s+1}$,
$g_0=16J/[U(2-\alpha_0)]$.
To summarize the low frequency form of the GLW action reads: 
\begin{eqnarray}
  \label{free1}\nonumber{}
  F[\bar{\psi},\psi]
  &=&\frac{1}{2\beta J N}
  \sum_{k,\omega_\mu}\left[\epsilon+\frac{k^2}{4}+\eta|\omega_\mu|^s 
    +\zeta\omega_\mu^2\right]\,|\psi(k,\omega_\mu)|^2\\
  & &+\kappa \int d^2r d\tau|\psi(r,\tau)|^4\,,
\end{eqnarray}
where $\epsilon=1-g_0$.  
Clearly the model exhibits a {\em non-ohmic}
dissipative dynamics, reducing to ohmic damping only in the special
case $s=1$.  Surprisingly, the ohmic damping in the quantum phase
model of Eq.~(\ref{Sjja}) yields a non-ohmic dynamics for the coarse-grained
order-parameter.  The quadratic frequency dependence dominates over
the dissipative dynamics for $s>2$, or equivalently $\alpha_0<2/3$.

The model exhibits a phase transition at $\epsilon(\alpha_0,J,U)=0$.
The ordered phase supports long-range superconducting order, whereas
the disordered phase is an insulator. The phase diagram for $T=0$ is
displayed in the inset of Fig.~1. Increasing damping shifts the phase
boundary to lower values of $J$.  The insulating phase disappears
for $\alpha_0>2$. In this region the dissipative processes completely
suppress the quantum fluctuations, in analogy with the dissipative
phase transition, discussed in Ref.~\cite{chak}.
\begin{figure}[htbp]
  \unitlength1cm
  \begin{picture}(8.3,6)
    \put(0,0.2){\psfig{figure=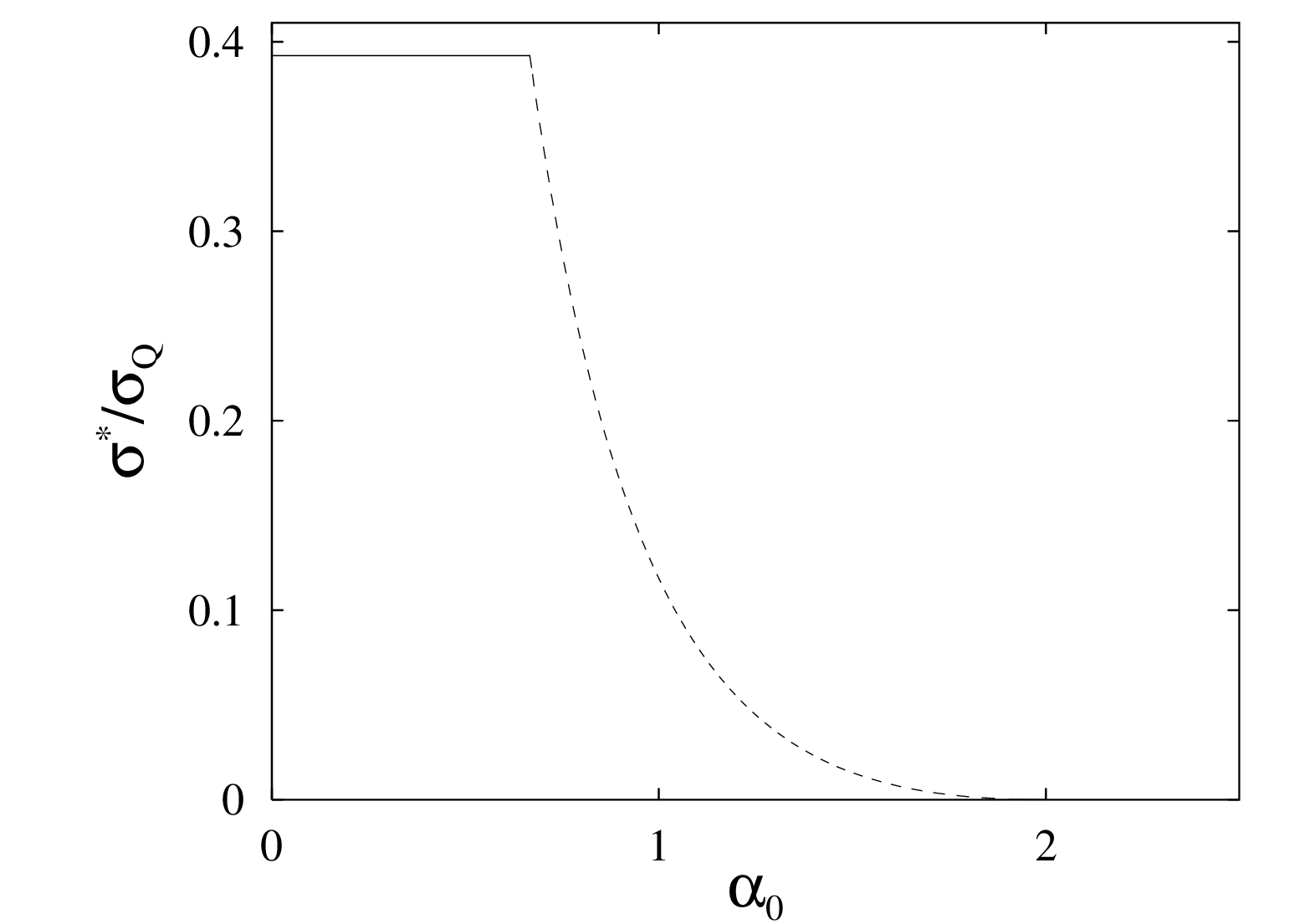,height=5.8cm,width=8cm}}
    \put(3,2.1){\psfig{figure=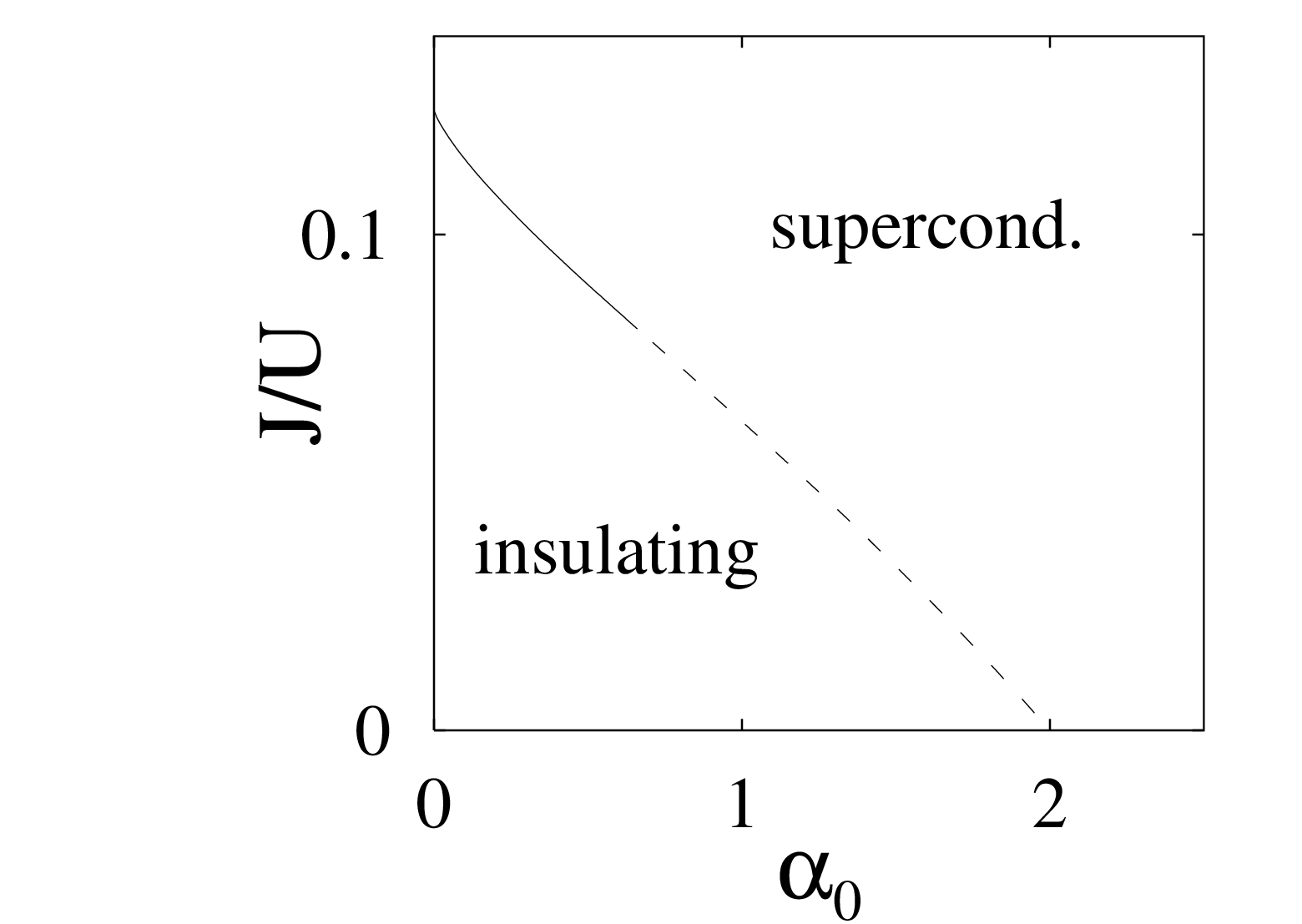,height=3.7cm,width=4.7cm}}
  \end{picture}
  \caption{Conductivity at the transition as a function of $\alpha_0$.
    The inset shows the corresponding phase diagram at $T=0$.  Along
    the solid line $\sigma$ is universal, whereas it is a function of
    $\alpha_0$ along the dotted line.}
\end{figure}

At $T=0$ a quantum phase transition takes place which is characterized
by a dynamical critical exponent $z$.  In the limit of weak damping,
the critical behaviour is that of the non-dissipative models, $z=1$.
In the general case it is given by $z=\text{max}(1,2/s)$, and the
critical behaviour is that of a $(2+z)$-dimensional XY-model.
Therefore our results establish the existence of a new
universality class of the superconductor - insulator transition for $s<2$.

This quantum phase transition attracted intense interest because of
the claim of a {\it universal} conductivity at 
criticality~\cite{fish1}.  To address the issue of universality we 
calculate the conductivity in terms of two and four point Green's
functions~\cite{cha1}.  In the Gaussian approximation the four point
function factorizes.  An analytic continuation
$i\omega_{\mu}\rightarrow\omega+i\delta$ yields the dependence on the
real-frequency~\cite{otte}
\begin{eqnarray}
  \label{freq} \nonumber{}  \sigma(\omega)&&
  =\frac{\sigma_Q}{16\pi\omega}\int_{-\infty}^{\infty}
  \frac{{d}z}{1-{e}^{-\beta z}}  \int_0^{\infty}{d}kk^3\\
  \nonumber{}
  &&\Big[G^{R}(k,z)-G^{A}(k,z)\Big]\times \Big[G^{R}(k,z)\\
  &&+G^{A}(k,z)-G^{R}(k,z+\omega)-G^{A}(k,z-\omega)\Big],
\end{eqnarray}
where $\sigma_Q=h/4e^2$. 
The advanced and retarded Greens functions are given by
\begin{eqnarray}
  \nonumber{}
  \Big[G^{A/R}(k,\omega)\Big]^{-1}&&=\epsilon+\frac{k^2}{4}-\zeta\omega^2\\
  &&+\eta|\omega|^s\Big[\cos\frac{s\pi}{2}
  \pm i\mbox{sign}(\omega)\sin\frac{s\pi}{2}\Big],
\end{eqnarray}
with $\epsilon\geq 0$. Without damping the frequency dependence
exhibits a gap of the size $\omega_0=\sqrt{4\epsilon/\zeta}$, thus this 
disordered phase is a Mott-insulator.
With increasing damping the Mott gap is smeared out.  For $s<2$ and
low frequencies $\omega\ll\omega_0=\sqrt{4\epsilon/\zeta}$ one finds
\begin{equation}
  \mbox{Re}\,\sigma(\omega)=\sigma_Q
  \frac{\eta^2\sin^2(\frac{\pi}{2}s)}{6\pi\epsilon^2} 
  \frac{[\Gamma(1+s)]^2}{\Gamma(2+2s)} |\omega|^{2s}\;.
\end{equation}
The conductivity shows a power-law behaviour at low frequency, where
the power depends on the strength of the dissipation for $s\leq 2$.
These two types of behaviours are displayed in Fig.~2.
\begin{figure}[htbp]
  \unitlength1cm
  \begin{picture}(8.3,6)
    \put(0,0.2){\psfig{figure=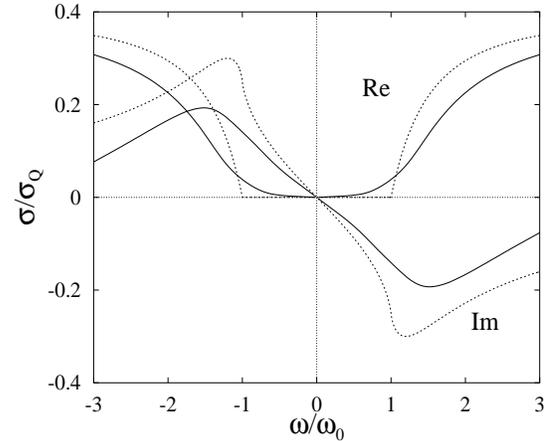,height=5.8cm,width=8cm}}
  \end{picture}
  \caption{Real and imaginary parts of the conductivity as a
    function of the frequency. Without dissipation $(\alpha_0=0)$ the
    real part exhibits a gap (dotted lines).  This gap is smeared in
    the presence of dissipation (solid lines). The damping parameters
    are $s=1/2$, and $\eta\omega_0^s/\epsilon=1$.}
\end{figure}
 
Of particular interest is the d.c.\ conductivity at the transition,
i.e.\ for $\omega_0 \rightarrow 0$ (with $\omega_0/\omega\rightarrow
0$).  As shown in Fig.~1 the value of the d.c.~conductivity at
criticality $\sigma^*$ is a function of the strength of the ohmic
damping for $\alpha_0 > 2/3$.  To summarize, the non-dissipative
superconductor - insulator transition possesses a finite basin of
attraction. In the region of $0\leq\alpha_0\leq2/3$ the dissipation is
an irrelevant operator.  It is characterized by $z=1$ and a universal
critical conductivity.  For strong enough coupling to localized
fermions $\alpha_0>2/3$ a new universality class is present, with
damping dependent $\sigma^*$ and $z$. This is the central result of
our paper.

Renewed interest in the superconductor - insulator transition was
generated by the early experimental report of a universal conductivity
at criticality, with the value
$(\sigma^*)^{-1}=6.5k\Omega$~\cite{orr}.  However, subsequent
measurements on superconducting films~\cite{liu,valles,yazd} and on
Josephson junction arrays~\cite{geer} reported resistivity values
distributed in the broad range of 2-20 $k\Omega$.  Our emphasis on the
electronic degrees of freedom offers a possible explanation of this
lack of universality.

In many experiments the transition is tuned by a magnetic field 
$\vec{B}$~\cite{paalanen,yazd}, which we therefore discuss next. 
The magnetic field is incorporated by introducing a vector potential
$\vec{A}(x,\tau)$ in Eq.~(\ref{free}). For weak frustration $f\ll 1$
the lattice structure can be neglected. The magnetic field leads to
the formation of Landau levels.  The longitudinal conductivity is
given by \cite{otte}
\begin{eqnarray}\nonumber
  \sigma(\omega_{\mu})=\sigma_Q\frac{\omega_c^2}{2\omega_\mu\beta}
  \sum_{n=0}^{\infty} \sum_{\omega_{\nu}} 
  (n+1)\Big[2 G_{\omega_{\nu},n} G_{\omega_{\nu},n+1}\\
  -G_{\omega_{\nu}+\omega_{\mu},n}G_{\omega_{\nu},n+1}
  -G_{\omega_{\nu},n}G_{\omega_{\nu}+\omega_{\mu},n+1}\Big]\;,
\end{eqnarray}
where $G_{\omega_{\mu},n}=
[\epsilon+\omega_c(n+1/2)+\eta|\omega_{\mu}|^s+\zeta\omega_{\mu}^2]^{-1}$,
with the dimensionless cyclotron frequency $\omega_c=\pi f$.
The magnetic frustration $f=\Phi/\Phi_0$ is proportional to the 
flux through a unit cell.
In the mean-field approximation the phase boundary is given by
$\epsilon+\omega_c/2=0$. The fact that the location of the phase boundary 
is magnetic field dependent demonstrates, that our formulation is capable
of capturing the field-tuned transition.

The analytic continuation to determine the conductivity as a function
of real frequencies follows the same lines as in the zero field case.
We explicitly evaluate the real part of the conductivity as shown in
Fig.~3.  The frequency dependence reflects the underlying Landau-level
structure. It is smeared due to the influence of the damping.
\begin{figure}[htbp]
  \unitlength1cm
  \begin{picture}(8.3,6)
    \put(0,0.2){\psfig{figure=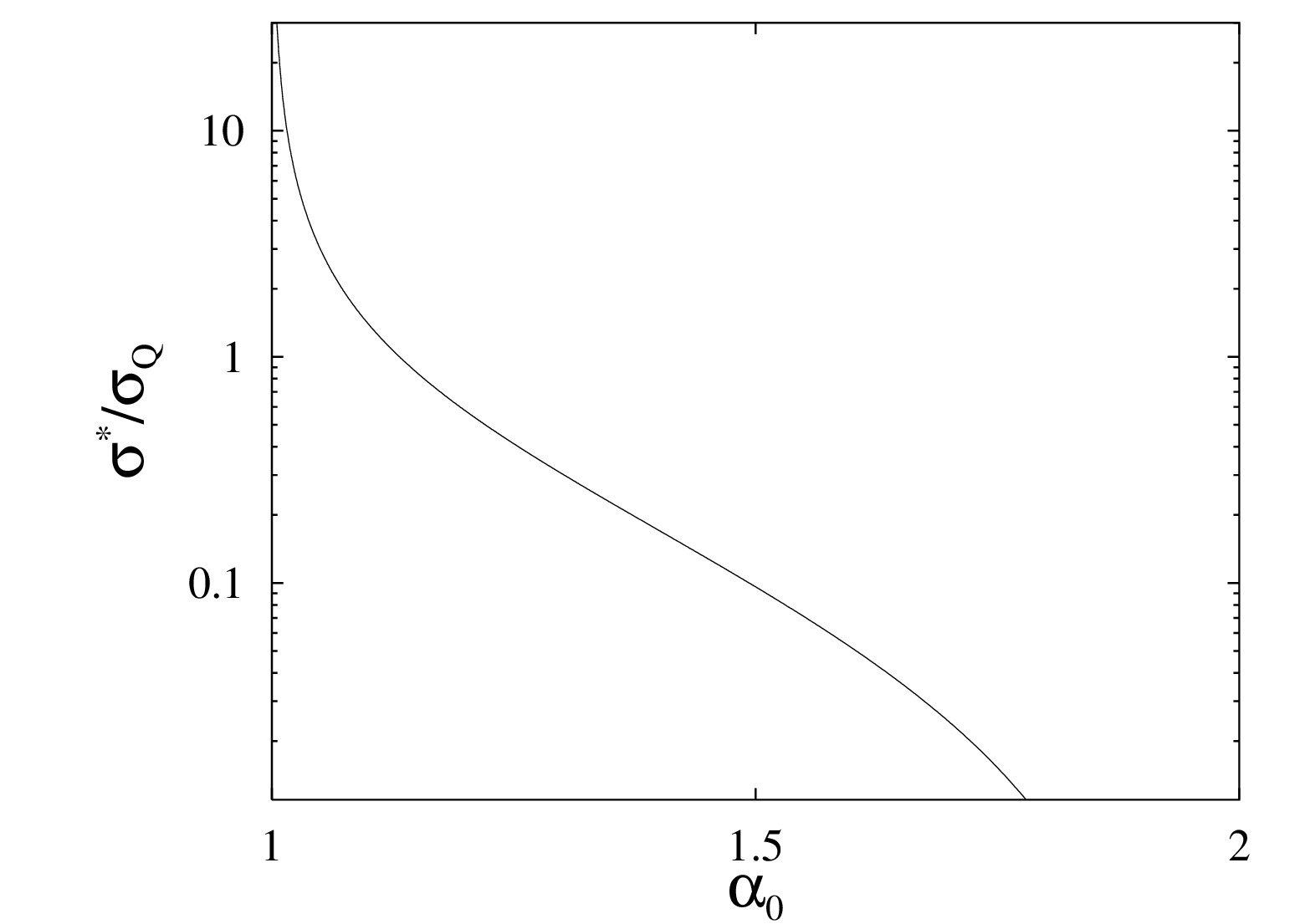,height=5.8cm,width=8cm}}
    \put(3.3,2.2){\psfig{figure=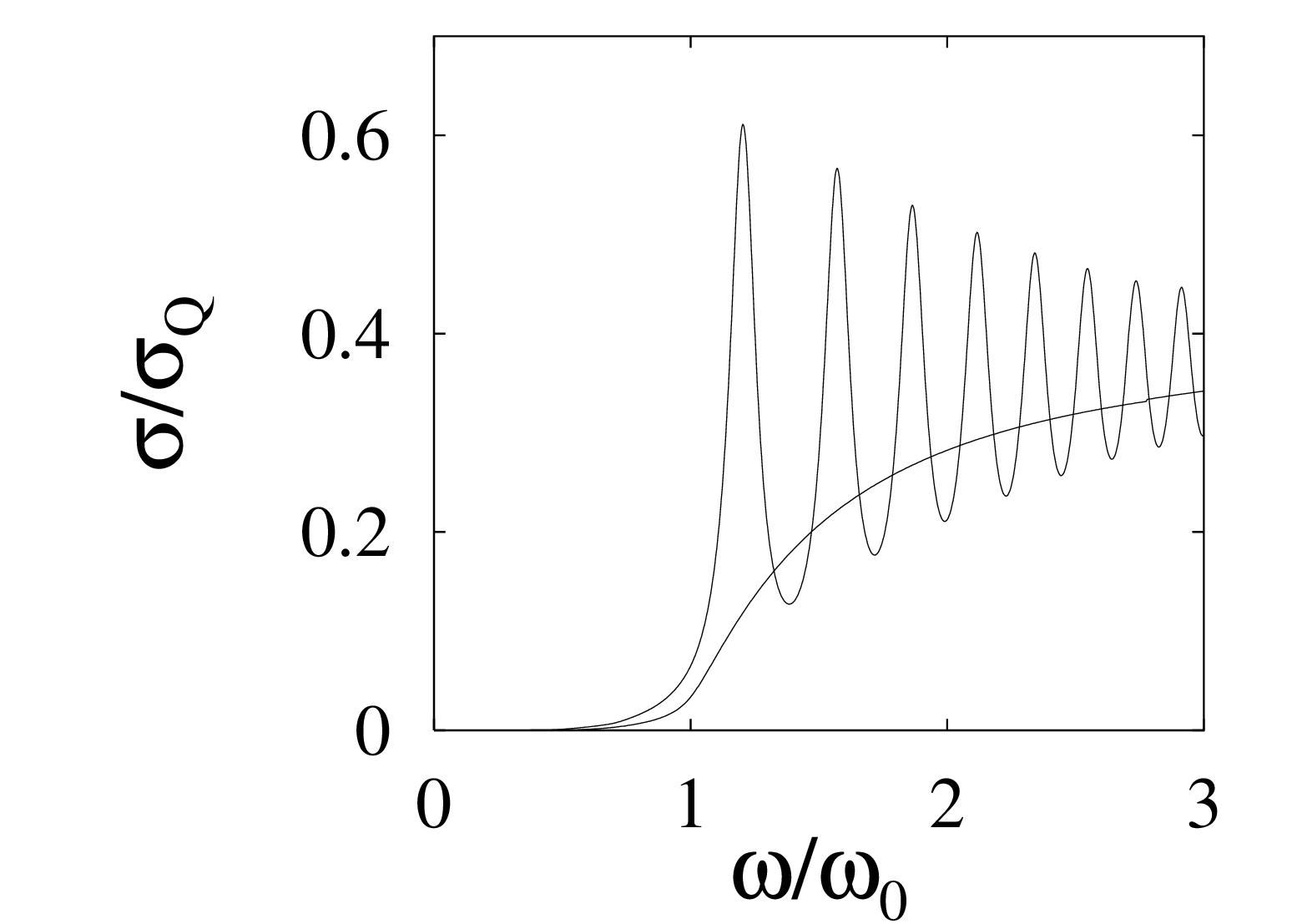,height=3.6cm,width=4.4cm}}
  \end{picture}
  \caption{Longitudinal conductivity at the transition in a finite magnetic 
    field as a function of $\alpha_0$.  The inset shows the real part
    of the frequency dependent longitudinal conductivity in a magnetic
    field (oscillating curve), compared to the zero field case (smooth
    curve). With the parameters
    $\omega_c/\protect{\sqrt{\epsilon}}=2$, $s=1$, and
    $\eta\omega_0/\epsilon=1/4$.}
\end{figure}
The d.c.~conducitvity {\it at} the transition is determined by the
lowest Landau level $n=0$. In the absence of dissipation it {\it
  diverges}.  Again we find that for strong enough damping,
$\alpha_0>1$, the critical conductivity $\sigma^*$ becomes {\it
  finite}, with the damping dependent value, as shown in Fig.~3.  For
$\alpha_0\rightarrow 1$ it assumes the asymptotic form 
$\sigma^*\sim 1/(\alpha_0-1)$.
 
The previous analysis showed that the effect of the local damping for
the phases has a great impact on the properties of the system. In the
second part of this paper we discuss various scenarios which can give
rise to local dissipation.

The key feature giving rise to dissipation is the presence of
gapless degrees of freedom.  These electronic degrees of freedom can
either be of intrinsic origin, or they can result from a coupling to
the substrate.  We argue that both scenarios are viable candidates for
the description of the experimental data.

Before reviewing the microscopic motivation we point out the important
difference between the local dissipation, investigated here, and the
non-local forms, studied in Refs.~\cite{chak}. When local damping is
present, the number of Cooper pairs is not conserved: they can decay
into a pool of normal electrons.  On the contrary, if the damping is
non-local, e.g.\ it is induced by inter-grain electron transfer, then
it conserves the charge within the film.  Because of this essential
difference it can be expected that genuinely new physics arises in the
present model.

In the remainder of the paper we discuss possible origins of the local
damping.  The two necessary ingredients for the proposed mechanism are
local pools of gapless electrons with a temperature independent
density, and a coupling term between these electrons and the phase of
the order parameter, which leads to ohmic dissipation.  
Once gapless electrons are present, the Andreev
process satisfies these criteria.  It describes the conversion
of normal electrons into Cooper pairs at superconductor - normal (SN)
interfaces.  In the absence of a tunnel barrier the charge transfer
across the interface is continuous. This results in a fully ohmic
dynamics for the superconducting phase. Therefore the proper
description of this process is exactly the dissipative term in
Eq.~(\ref{Sjja}).  For the case of two superconducting islands,
connected by a normal metal, a microscopic derivation of the ohmic dynamics
was provided in Ref.~\cite{schoen}. Adapting their derivation
for the present case of an SN structure confirms our statement
explicitly.  In the ideal case the value of the Andreev resistance is half 
that of the normal state resistance, consequently the arising local 
dissipation can be large.  We emphasize that this process involves pools 
of {\it  localized} electrons which are therefore not part of the conducting
path.  Transport across the sample is dominated by tunneling between
grains, resulting in a much higher subgap resistance.

The remaining question is the origin of the gapless electrons.  In
order to give rise to ohmic dynamics, the level spacing of their
spectrum has to be small compared to all other energy scales in the
problem.  This necessitates a relatively large spatial extent of their
wavefunction, which can be estimated to be in the range of hundreds of
{\AA}ngstroms.  An intrinsic scenario can be suggested by recalling
that in the absence of superconductivity the system is an insulator,
otherwise the experiments would display a superconductor - metal
transition.  On the other hand, in order to be able to support
superconductivity, it has to be close to the metal - insulator
transition. In this region the combined effects of disorder and
enhanced Coulomb repulsion act as very effective mechanisms to break
up the Cooper pairs into normal electrons~\cite{finkel}.  In the
vicinity of the transition the localization length of the electrons is
strongly enhanced. Thus their spatial extent is large, leading to a
considerable density of states at the Fermi energy.  The existence of
such low energy excitations with a density comparable to the normal
state density was observed in tunneling experiments~\cite{valles}.

An extrinsic origin of the local damping mechanism can be due to a
conducting substrate. In this case the unpaired electrons of the
ground plane constitute the reservoir.  It is well established that in
proximity coupled junction arrays the metallic substrate indeed is a
source of localized dissipation~\cite{theron,shaw,korshunov}.
Superconducting films are commonly grown on disordered semiconductor
substrates.  As was suggested in Ref.~\cite{liu} the semiconductor
substrate might be able to support metallic transport on short length
scales. It is possible that the deposited metal dopes the amorphous
semiconductor substrate, giving rise to a temperature independent density
of states at the Fermi level. An additional remarkable possibility was
raised in connection to the study of Ag/Ge, where the metal -
semiconductor interface {\it itself} seems to support metallic
transport~\cite{liu2}.

In sum we developed a modified understanding of the superconductor -
insulator transition. We emphasized the importance of
localized gapless electronic excitations. They give rise to a local
ohmic dissipation for the phase of the superconducting order
parameter. A new universality class of the transition was identified
for strong enough dissipation.  It is characterized by a {\it
  non-universal} value of the critical conductivity and a damping
dependent dynamical critical exponent. The dependence of the above
results on a magnetic field was also analyzed.  We reviewed several
possible microscopic origins of the gapless electrons, and identified
the Andreev process as a likely source of the ohmic dissipation.

We would like to thank Rosario Fazio and Arno Kampf for valuable
discussions. This work is within the research program of SFB 195 and
supported by the DFG. GTZ was also supported by the grant NSF DMR
95-28535, and AvO by the Swiss National Fonds.

\end{document}